\documentclass[referee,sn-basic]{sn-jnl}

\DeclareUnicodeCharacter{02BC}{'}

\usepackage{graphicx}%
\usepackage{multirow}%
\usepackage{amsmath,amssymb,amsfonts}%
\usepackage{amsthm}%
\usepackage{mathrsfs}%
\usepackage[title]{appendix}%
\usepackage{xcolor}%
\usepackage{textcomp}%
\usepackage{manyfoot}%
\usepackage{booktabs}%
\usepackage{algorithm}%
\usepackage{algorithmicx}%
\usepackage{algpseudocode}%
\usepackage{listings}%
\usepackage[version=4]{mhchem} 
\usepackage[]{lineno}

\begin{document}

\title[Article Title]{Spin crossover in FeO under shock compression}


\author*[1,2]{\fnm{L{\'e}lia} \sur{Libon}}\email{llibon@carnegiescience.edu}

\author*[3]{\fnm{Alessandra} \sur{Ravasio}}\email{alessandra.ravasio@polytechnique.edu}

\author[1]{\fnm{Silvia} \sur{Pandolfi}}

\author[4]{\fnm{Yanyao} \sur{Zhang}}

\author[5]{\fnm{Xuehui} \sur{Wei}}

\author[6]{\fnm{Jean-Alexis} \sur{Hernandez}}

\author[4,5]{\fnm{Hong} \sur{Yang}}

\author[4]{\fnm{Amanda J.} \sur{Chen}}

\author[3]{\fnm{Tommaso} \sur{Vinci}}

\author[3]{\fnm{Alessandra} \sur{Benuzzi-Mounaix}}

\author[7]{\fnm{Clemens} \sur{Prescher}}

\author[8]{\fnm{François} \sur{Soubiran}}

\author[9]{\fnm{Hae Ja} \sur{Lee}}

\author[9]{\fnm{Eric} \sur{Galtier}}

\author[9]{\fnm{Nick} \sur{Czapla}}

\author[4,9]{\fnm{Wendy L.} \sur{Mao}}

\author[4,9]{\fnm{Arianna E.} \sur{Gleason}}

\author[5]{\fnm{Sang Heon} \sur{Shim}}

\author[9]{\fnm{Roberto } \sur{Alonso-Mori}}\email{robertoa@slac.stanford.edu}

\author*[1,2]{\fnm{Guillaume} \sur{Morard}}\email{ guillaume.morard@cnrs.fr}

\affil*[1]{\orgdiv{Institut de Minéralogie, de Physique des Matériaux et de Cosmochimie}, \orgname{Sorbonne Université, Muséum National dʼHistoire Naturelle, UMR CNRS 7590}, \orgaddress{\street{4 Place Jussieu}, \city{Paris}, \postcode{75005}, \country{France}}}

\affil[2]{\orgdiv{ISTerre}, \orgname{Université Grenoble Alpes, CNRS}, \orgaddress{\city{Grenoble}, \postcode{38000}, \country{France}}}

\affil[3]{\orgdiv{Laboratoire pour l’Utilisation des Lasers Intenses (LULI)}, \orgname{Ecole Polytechnique, CNRS, CEA, Sorbonne Université}, \orgaddress{\city{Palaiseau}, \postcode{91128}, \country{France}}}

\affil[4]{\orgdiv{Earth and Planetary Sciences}, \orgname{Stanford University}, \orgaddress{ \city{Stanford}, \postcode{94305}, \state{CA}, \country{USA}}}

\affil[5]{\orgdiv{School of Earth and Space Exploration}, \orgname{Arizona State University}, \orgaddress{\city{Tempe}, \postcode{85287}, \state{AZ}, \country{USA}}}

\affil[6]{\orgname{European Synchrotron Radiation Facility}, \orgaddress{\city{Grenoble}, \country{France}}}

\affil[7]{\orgdiv{ Institut für Geo- und Umweltnaturwissenschaften}, \orgname{Albert-Ludwigs-Universität
Freiburg}, \orgaddress{\city{Freiburg}, \postcode{79104}, \country{Germany}}}

\affil[8]{\orgname{CEA/DAM Ile-de-France}, \orgaddress{\city{Arpajon}, \postcode{91297}, \country{France}}}

\affil[9]{\orgdiv{SLAC National Accelerator Laboratory} \orgaddress{\street{2575 Sand Hill Rd.}, \city{Menlo Park}, \postcode{94025,}, \state{CA}, \country{USA}}}


\abstract{FeO (wüstite), which exhibits complex electronic and structural properties with increasing pressure and temperature, is a key mineralogical phase for understanding deep planetary interiors. However, direct measurements of its spin state at high-pressure and temperature remain challenging in static compression experiments. Here, we employ laser-driven shock compression to extend the FeO principal Hugoniot up to $\sim$900~GPa and perform in situ X-ray diffraction and X-ray emission spectroscopy up to 250~GPa, probing FeO’s crystal structure and spin state. We demonstrate a continuous spin crossover of iron in FeO over a broad pressure range, with the high-spin state persisting beyond Earth’s core–mantle boundary (CMB) conditions. These observations provide new experimental constraints on iron spin state at extreme conditions essential for geophysical models of (exo)planetary interiors.}



\maketitle


FeO (\ce{Fe_{x}O}, 0.9$<$x$<$0.98), also known as wüstite, is a key phase in the Earth's mantle and core. Its physical and chemical behavior under extreme pressure-temperature conditions has been the subject of extensive research due to its significant implications for understanding the structure and evolution of the Earth and for Earth-like planets \citep{Coppari2021, Mao1996, Hazen1984}. It represents the iron-rich end-member of ferropericlase \ce{(Mg,Fe)O}, the second most abundant mineral in the lower mantle \citep{Morard2017,McCammon1983}, and is believed to occur in iron-enriched layers near the core–mantle boundary (CMB) due to partial melting or chemical segregation. Experimental and theoretical studies further suggest that an FeO-rich layer may also have formed through the fractional crystallization of a primordial basal magma ocean \citep{Boukare2025,Boukare2015} or through core-mantle chemical interactions \citep{Hirose2017, manga1996FeO, Knittle1991CMB}. Additionally, seismic observations, combined with experimental constraints, suggest that ultra-low velocity zones (ULVZs) can be explained by highly FeO-rich solid materials \citep{Jackson2021,Dobrosavljevic2019,Wicks2017,Wicks2010,Ohta2014,Nomura2011,Labrosse2007}. FeO is also central to understand the core composition, as oxygen is one of the main light elements in the core \citep{Komabayashi2014,Badro2007,Alfe2002,Alfe1999,Ringwood1977}. 
Thus, its physical properties at extreme conditions have significant implications for the structure, dynamics, and long-term evolution of planetary interiors.

Despite being a simple binary oxide, FeO exhibits a complex phase diagram, undergoing multiple pressure-induced structural transitions from NaCl-type B1 structure (F\textit{m}$\overline{3}$\textit{m}) to rhombohedral rB1 phase (R$\overline{3}$\textit{m}) at 16-27~GPa at ambient temperature \citep{Jacobsen2005,Yagi1985}, to the NiAs-type B8 phase (P6$_3$/\textit{mmc}) $>$70~GPa \citep{Murakami2004,Mao1996,Ozawa2010, Fischer2011EOS}. At $>$250~GPa and $>$2000~K, FeO adopts the CsCl-type cubic B2 structure \citep{Coppari2021,Ozawa2011Stratification}. More recently, \citet{LiX2024} reported a monoclinic mP4 phase (P2$_1$/\textit{m} space group) at around $\sim$80~GPa on quenched sample after heating, revealing further complexity to the high-pressure diagram of FeO.

FeO also exhibits pressure-induced electronic transitions. First, an insulator-to-metal transition (IMT) in the B1 phase has been experimentally investigated at $\sim$30-90~GPa and $\sim$2600-1400~K \citep{Fischer2011Metallization, Ohta2012}. Recently, however, it has been argued that instead of a sharp transition, FeO-B1 enters a broad quantum critical state between its insulating and metallic states \citep{Ho2024}, which could account for the electrical conductivity changes observed experimentally \citep{Ohta2012}. Second, FeO undergoes a spin transition from high-spin (HS) to low-spin (LS) iron \citep{Hamada2016,Ozawa2011Stratification,Ozawa2011Spin,Badro1999,Pasternak1997}. In the solid state, this transition can lead to a 20 to 40~\% reduction in the ionic volume of iron and affect FeO’s compressibility, density, and bulk modulus \citep{Greenberg2023,Pasternak1997}. In the liquid state, LS Fe–O bonds are shorter than in the HS state, resulting in a denser liquid relative to its HS equivalent \citep{Morard2022,Hamada2016}.


Additionally, LS iron appears to preferentially partition into the liquid phase, enhancing iron diffusion, particularly within ULVZs \citep{Lin2013,manga1996FeO}, which may contribute to the stabilization of a dense, Fe-rich liquid layer at the core–mantle boundary (CMB) within a basal magma ocean \citep{Labrosse2007}. These processes have major implications for the long-term evolution of Earth’s deep interior, influencing mantle convection, core dynamics, and chemical exchange across the CMB. Thus, constraining the spin state of iron and its depth-dependent evolution is essential for developing accurate models of Earth’s inner structure and evolution. 

Previous studies have suggested that the iron spin transition is directly coupled to the IMT transition \citep{Leonov2015,Ohta2012,Cohen1997}. Following the recent work of \citet{Ho2024}, the relationship between these two transitions depends on temperature: at lower temperatures, the IMT can proceed via closure of the t$_{2g}$ gap prior to spin crossover, whereas at higher temperatures, the metallization occurs through the closure of the e$_g$ gap, which simultaneously triggers the spin crossover, implying that spin and electronic transitions may become partially coupled. In contrast, recent density-functional theory+dynamical mean-field theory (DFT+DMFT) calculations, supported by in situ X-ray diffraction (XRD) and X-ray emission spectroscopy (XES) measurements in laser-heated diamond anvil cells (LH-DACs), have challenged this assumption \citep{Greenberg2023}. Those experiments report instead a gradual HS to a LS crossover within metallic FeO-B1, suggesting a decoupling between the two transitions even at high temperature. While such LH-DAC studies represent major advances in directly probing FeO's spin state at high P-T conditions, they also highlight constraining factors that limit the number of data acquired and make it difficult to fully track the spin crossover experimentally, particularly at high temperatures. These contrasting results motivate a complementary experimental approach capable of extending the pressure-temperature range to fully probe the spin crossover.


Here, we employ dynamic laser-driven shock compression to access high pressure-temperature conditions along the principal Hugoniot that extend well beyond the P-T range reached in static compression experiments. We report results from two experimental campaigns on FeO. First, we extended the Hugoniot curve of FeO up to $\sim$900~GPa. Secondly, using in situ XRD and XES, we directly tracked both the structural and the iron-spin state evolution up to 261~GPa along the Hugoniot. Our experimental investigations were conducted on a \ce{Fe_{0.90}O} sample—referred to as FeO hereafter (details in supplementary material). Experiments were performed at the LULI2000 laser facility of the \textit{Laboratoire pour l’Utilisation des Lasers Intenses} (Institut Polytechnique de Paris, France) and at the Matter in Extreme Conditions (MEC) end-station \citep{Glenzer2016,Nagler2015} of the Linac Coherent Light Source (LCLS) at SLAC National Accelerator Laboratory.

\section*{Results and discussion}

\subsection*{Hugoniot measurements}

\begin{figure}[b]
\centering
\includegraphics[width=0.55\textwidth]{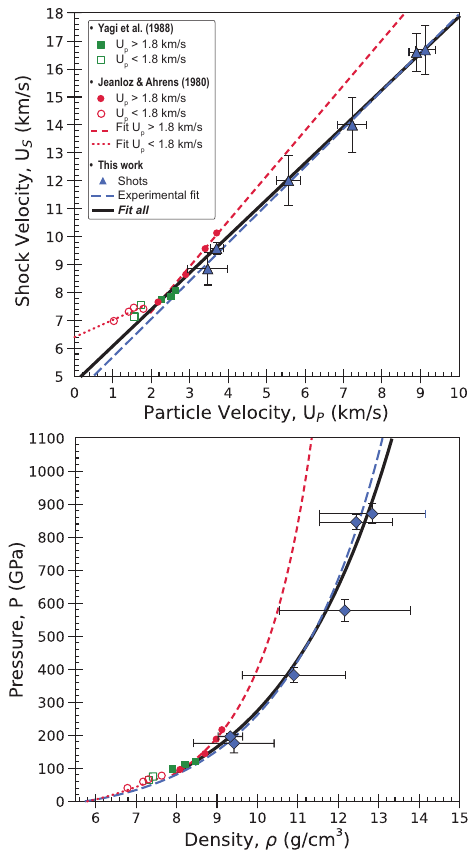}
\caption{FeO principal Hugoniot relationship in (left): shock velocity (U$_{S}$)-particle velocity (U$_{P}$) relationship and (right) pressure-density (P-$\rho$) relationship. Experimental data from this work (blue triangle) are fitted together with \citet{Jeanloz1980} (red circle), and \citet{Yagi1988} (green square) data from U$_{P}$ $>$ 1.8 km/s (black line). The red and blue dashed lines are the fit \citet{Jeanloz1980}'s data only and the data from this study only, respectively}\label{Fig-Hugoniot}
\end{figure}

We extended the FeO Hugoniot to near-terapascal pressures by performing 5 laser-driven shock compression experiments at LULI200, spanning from 176~GPa to 871~GPa (Fig.~\ref{Fig-Hugoniot}, Table~\ref{TAB-Results Hugoniot} and S3). This significantly expands the range of previous Hugoniot measurements \citep{Yagi1988, Jeanloz1980} to near-terapascal pressures. To determine the compression state in the FeO sample, we used the impedance matching technique \citep{Brygoo2015, zeldovich1966}. The $\alpha$-quartz window and aluminum pusher layer were used as reference materials for the impedance matching analysis. The $\alpha$-quartz served as a transparent witness, which enable direct measurements of the shock velocity, allowing to perform the impedance matching at the both the Al-quartz and Al-FeO interfaces. Details on the experimental configurations and data analysis procedure are provided in the Methods and Supplementary Materials. 

\begin{table}[h]
\caption{Hugoniot result for \ce{Fe_{x}O} (x = 90) including the target thickness (D), the shock velocity (Us), particle velocity (Up), pressure (P), and density ($\rho$). The FeO’s initial density of this study is $\rho$$_{0}$=5.73 g/cm$^{3}$.} \label{TAB-Results Hugoniot}
\begin{tabular}{@{}cccccccc@{}}
\toprule
Shot n° &  D$^{FeO}$ & U$_{S}$$^{quartz}$& U$_{S}$$^{FeO}$ & U$_{P}$$^{FeO}$ & P$_{H}$$^{FeO}$ & $\rho$$^{FeO}$ \\
 & ($\mu$m) & (km/s)& (km/s) & (km/s) & (GPa) & (g/cm$^{3}$) \\
\midrule
1 &  70 & 19.40 $\pm$ 0.17 & 16.69 $\pm$ 0.87 & 9.13 $\pm$ 0.26 & 871 $\pm$ 30 & 12.88 $\pm$ 1.3  \\
2 & 67 & 19.10 $\pm$ 0.12 &  16.60 $\pm$ 0.68 & 8.90 $\pm$ 0.20 & 846 $\pm$ 22 & 12.44 $\pm$ 0.91 \\
35 & 63 & 13.33 $\pm$ 0.42  &  12.01 $\pm$ 0.90 & 5.57 $\pm$ 0.32 & 382 $\pm$ 23 & 10.90 $\pm$ 1.27 \\
40 & 64 & 16.25 $\pm$ 0.38 & 14.00 $\pm$ 0.97 & 7.23 $\pm$ 0.38 & 578 $\pm$ 33 & 12.16 $\pm$ 1.63 \\
73\footnotemark[1] & 65 & - & 8.85 $\pm$ 0.58 & 3.47 $\pm$ 0.52 \footnotemark[1]& 176 $\pm$ 29 & 9.42 $\pm$ 0.99 \\   
279 & 43 & - &9.58 $\pm$ 0.22 & 3.70 $\pm$ 0.17\footnotemark[1] & 196$\pm$10 & 9.33$\pm$0.3 \\
\botrule
\end{tabular}
\footnotetext[1]{from FeO free-surface measurement}
\end{table}

Measurements of U$_{P}$-U$_{S}$ (particle velocity–shock velocity) and $\rho$-P (density-pressure) for shocked FeO are presented in Fig.~\ref{Fig-Hugoniot}, together with previously published data \citep{Yagi1988, Jeanloz1980}. The initial density is a critical parameter in determining a material's Hugoniot. Our FeO sample has a different stoichiometry compared to previous work (\citet{Jeanloz1980}: x=0.94; \citet{Yagi1988}: x=0.91 and 0.95; and this study: x=0.90). In addition, to account for sample porosity in \citet{Jeanloz1980}, we recalculated the initial and shocked densities as detailed in the Supplementary Materials, section 2.2 and Table S1. Previous studies provide no evidence that small stoichiometric differences significantly affect the Hugoniot. Additionally, the discontinuous density increase near 70~GPa reported in earlier studies \citep{Yagi1988, Jeanloz1980} was not reproduced in static compression experiments. Although the origin of this discontinuity has been debated and attributed to various processes \citep{Jackson1990,Jackson1981}, including a phase transition \citep{Fei1994,Jeanloz1980} or metallization \citep{Fischer2011Metallization, Knittle1986Metallization,Knittle1986}, we follow previous work \citet{Young2021} and consider only data above this discontinuity (P $>$73~GPa; U$_{P}$ $>$1.8~km/s and $\rho$ $>$7.8~g/cm$^3$), since our lowest measured pressure exceeds this threshold (82~GPa; see Table~S2).

In Fig.~\ref{Fig-Hugoniot}, the measured U$_{P}$-U$_{S}$ and $\rho$-P relationship are compared with the earlier gas-gun experiments and its high-pressure extrapolation \citep{Yagi1988,Jeanloz1980}. In the limited range where the datasets overlap (3.5~$<$~U$_{P}$~$<$ 4~km/s, corresponding to P $\sim$ 200~GPa), our measurements are in good agreement with the gas-gun results. At higher pressures, however, our measurements diverge from a simple extrapolation of the gas-gun results, underscoring the importance of direct experimental constraints. Our laser-driven shock experiments extend to substantially higher-pressure and reveal lower shock velocities at a given particle velocity, defining a higher-density Hugoniot than previously projected. These results highlight the need of our direct high-pressure measurements to robustly constrain the FeO Hugoniot.

Importantly, shot \#279 provides an absolute equation of state measurement, for which both the U$_{P}$ and the U$_{S}$ were independently determined: U$_{P}$ from the FeO free-surface velocity using the approximation  U$_{fs}$ = 2$_{P}$, valid at moderate pressures \citep{zeldovich1966} and U$_{S}$ from the shock transit time, respectively. This enables a direct determination of the FeO pressure using the Rankine-Hugoniot relations without relying on the impedance matching. This data point is in good agreement with both the other measurement from this study, and the previously published gas-gun data, validating the high-pressure Hugoniot path reported here.

Combining our results with the lower pressure data, the FeO Hugoniot is described by a U$_{P}$-U$_{S}$ relationship: U$_s$= s $*$ U$_p$ + C$_0$ with s=~1.3089~($\pm$0.12) and C$_0$=~4.7775~($\pm$0.59), for U$_p$~$\geqslant$1.8~km/s.

\subsection*{X-ray measurements}
In situ X-ray diffraction (XRD) and X-ray emission spectroscopy (XES) measurements during the shock compression were performed at the Matter in Extreme Conditions (MEC) end-station of the Linear Coherent Linac Source (LCLS) at SLAC National Accelerator Laboratory. This configuration, recently applied to dense silicate \citep{Shim2023}, enabled us to track both structural changes and spin state evolution in situ.

\begin{figure}[b]
\centering
\includegraphics[width=0.45\textwidth]{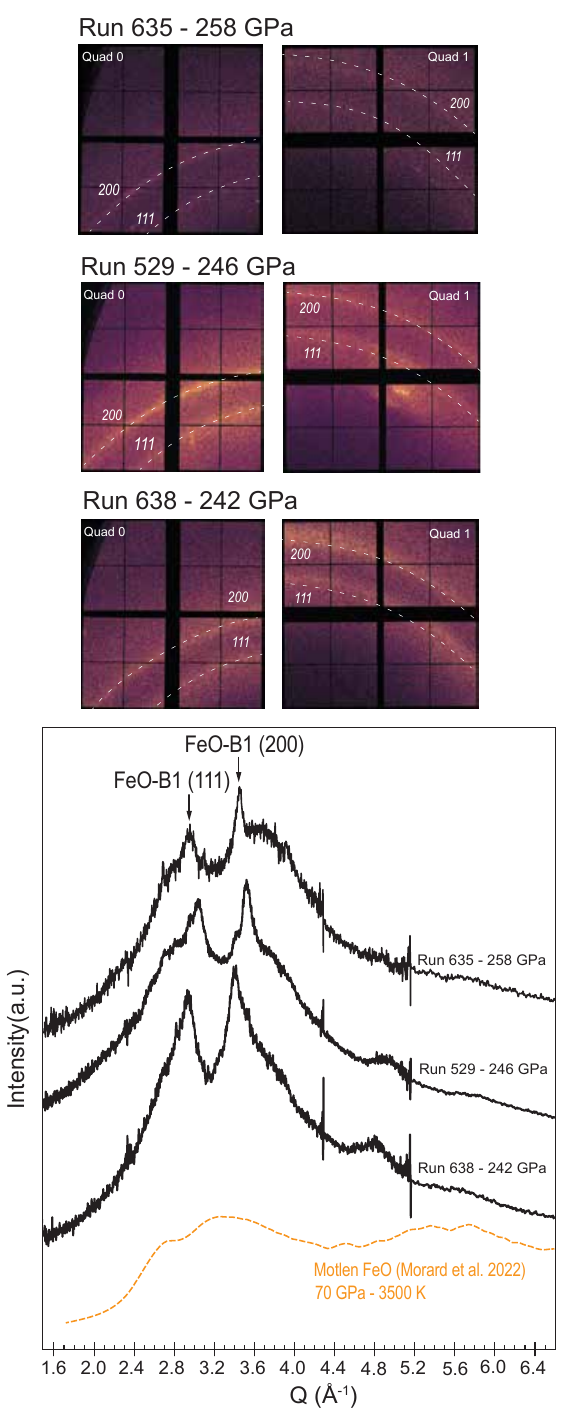}
\caption{X-ray diffraction images from Q0 (2$\theta$: 15° to 35° and Q-range: 22 to 50 nm$^{-1}$) and Q1 (2$\theta$: 19° to 28° and Q-range: 13 to 40 nm$^{-1}$) from ePix 10k detectors and 2D integrated patterns from Q0, Q1, and Q2 at 17 keV for run 638 - 242~GPa, run 529 - 246~GPa, and run 635 - 258~GPa. The orange dashed line is the FeO liquid structure factor from \citet{Morard2022}}\label{Fig-XRD-XES}
\end{figure}

Simultaneous XRD and XES measurements could be performed on shocked FeO at 9~keV. To extend the Q-range of the XRD, additional shots (runs 529, 635, and 638; see Table S2) were acquired at 17~keV X-ray energy (Fig.~\ref{Fig-XRD-XES}), though at the expense of XES. The 9~keV XRD data were inconclusive due to self-absorption from the unshocked sample layer. Indeed, to ensure accurate XES measurements, the signal was collected from the drive laser side at a 60° angle. In this configuration, the XFEL probe was timed so that the emission signal was acquired within the first $\sim$20~$\mu$m of shock propagation into the sample. This ensured a homogeneous compression state and minimized release effects, but with total sample thicknesses ranging from $>$30 to 50~$\mu$m, the transmitted XRD signal had to pass through an additional 10–30~$\mu$m of unshocked FeO before reaching the detector. Since the attenuation length of FeO is only $\sim$10~$\mu$m at 9~keV (compared to $\sim$40~$\mu$m at 17~keV), the self-absorption attenuated significantly the 9~keV XRD signal, preventing reliable interpretation of the solid phase's signal. 

In contrast, XRD at 17~keV clearly revealed the persistence of the high-pressure FeO-B1 structure at $>$242~GPa along the principal Hugoniot (Fig. \ref{Fig-XRD-XES}). Alongside the solid B1 peaks, two broad diffuse scattering features at Q = 2.93~$\mathring{A}$$^{-1}$ and Q = 3.4~$\mathring{A}$$^{-1}$ are observed (Fig.~\ref{Fig-XRD-XES}), consistent with the loss of long-range order. Such diffuse scattering may reflect either melting or rapid amorphization under nanosecond compression, with no evidence for a structural transition to the B2 phase \citep{Coppari2021, Fischer2011EOS,Ozawa2011Stratification,Murakami2004,Mao1996}. Because diffusion and segregation require longer timescales than the nanosecond timescales in laser-driven shocks, metastable or amorphous states can persist beyond their equilibrium stability fields, potentially explaining the survival of B1 peaks above 242 GPa. This kinetic effect, together with the lack of direct shock temperature measurements makes detection of melting along the Hugoniot challenging, as the loss of Bragg peaks and the emergence of diffuse scattering can also arise from amorphization, especially in crystalline samples. 

Despite these uncertainties, the persistence of FeO-B1 up to 258~GPa occurs well above the pressures where the previous FeO Hugoniots reported by \citet{Jeanloz1980} intersect published melting curves (between 140 and 180~GPa)
\citep{Fu2024,Dobrosavljevic2023,Morard2020,Komabayashi2014,Fischer2010}. Our experimental configuration does not allow for direct shock temperature measurements, and differences in starting material stoichiometry and initial density are expected to affect Hugoniot temperatures. Moreover, given the significant divergence between the high-pressure extrapolation of \citet{Jeanloz1980} Hugoniot and the Hugoniot measured in this study (Fig. \ref{Fig-Hugoniot}) makes it not straightforward to apply their temperature estimates, especially when extrapolated beyond the experimental range. 

Nevertheless, the diffuse scattering features observed above 242~GPa resemble those observed in static compression experiments of molten FeO (Fig.\ref{Fig-XRD-XES})\citep{Morard2022}, supporting a melting interpretation. We therefore suggest that FeO along the Hugoniot has a regime of solid-liquid coexistence above 240 GPa, although a contribution from hot amorphous states cannot be excluded. 

\begin{figure}[b]
\centering
\includegraphics[width=0.6\textwidth]{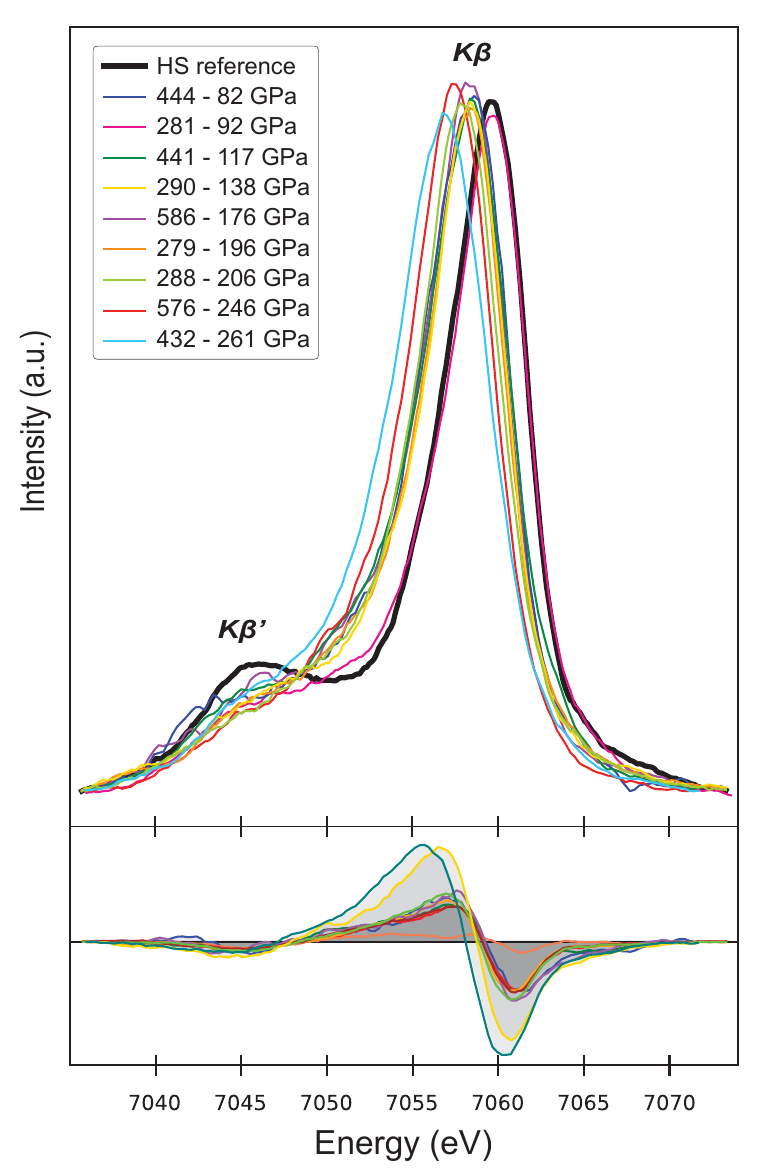}
\caption{\textbf{Top:} Fe K$\beta$ X-ray emission spectra collected from shocked FeO between 82 and 261~GPa. The black curve shows the HS reference spectrum at ambient pressure. \textbf{Bottom:} Difference curves between the reference and shocked spectra. The IAD (Integrated Absolute Difference) value \citep{Vanko2006} corresponds to the integral of the absolute difference of the solid gray areas. Detailed thermodynamic conditions and IAD values for each shot are listed in Table S2 of the Supplementary Material. Separated XES spectra of all shots are shown in Fig.~S4.}
\label{Fig-XES}
\end{figure}

In situ XES enabled direct probing of the iron spin state under shock compression. The HS reference spectrum at ambient conditions shows a main K$\beta$$_{1,3}$ peak and a lower-energy satellite peak (K$\beta$$^{\prime}$) (Fig.~\ref{Fig-XES}). Upon compression, the K$\beta$$_{1,3}$ peak systematically shift toward lower energies, while the K$\beta$$^{\prime}$ peak intensity progressively decreases (Fig.~\ref{Fig-XES}).

\begin{figure}[b]
\centering
\includegraphics[width=0.9\textwidth]{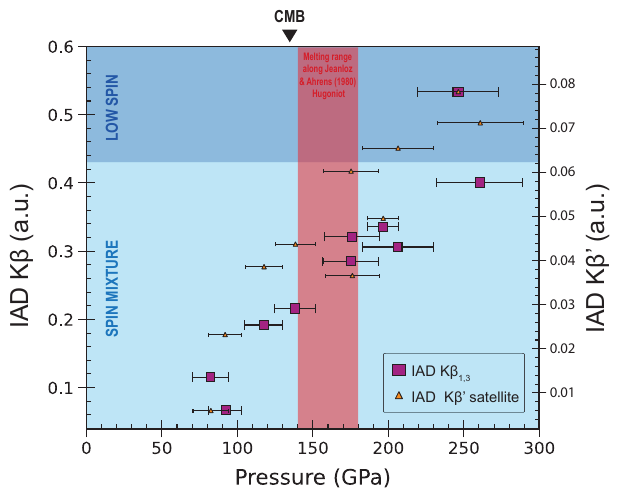}
\caption{Estimated IAD values \citep{Vanko2006} as a function of pressure for shocked FeO. Estimated LS IAD values are indicated by the darker blue area from K$\beta$$^{\prime}$ peak disappearance. Red areas indicate the estimated pressure where the melting curves \citep{Fu2024,Fischer2010,Morard2020,Komabayashi2014} crosses \citet{Jeanloz1980}'s Hugoniot. }\label{Fig-spin crossover}
\end{figure}

To estimate the spin state evolution, we applied the Integrated Absolute Difference (IAD) method \citep{Vanko2006}. In the absence of a well-characterized LS FeO reference spectrum under comparable conditions, we refrained from using a LS reference from another Fe-bearing mineral, as the difference in host chemistry and Fe valence state could introduce large uncertainties. Instead, we adopted a relative approach using the same HS reference spectrum for each experimental campaign to ensure consistency (Fig.~\ref{Fig-XES}). For each spectrum, the HS reference was subtracted from the shocked spectrum, and the absolute value of the difference curve was integrated over the energy range, resulting in the IAD value. A larger IAD value—larger shaded areas—represents a greater difference from the HS reference and hence a higher LS fraction. The observed spectral evolution (Fig.~\ref{Fig-XES}), including the disappearance of K$\beta$$^{\prime}$ peak and the systematic shift of the K$\beta$$_{1,3}$ main peak to lower energy, together with the continuous rise in IAD values with pressure (Fig.~\ref{Fig-spin crossover}), indicates a continuous spin crossover along the Hugoniot up to 200-250~GPa. 

Previous work by \citet{Albers2022a} has shown that pressure-induced structural distortion can shift and broaden the main K$\beta$$_{1,3}$ line, thereby affecting the IAD. In such cases, the IAD variation is not directly linked to spin transition but rather to changes in Fe-O bonding geometry/angles and in covalency. Consequently, the full-spectrum IAD (IAD K$\beta$$_{1,3}$ in Fig.~\ref{Fig-spin crossover}) may include non-spin-related contributions. This effect is especially true if there is a structural transition, such as the B1 to rB1 distortion observed in FeO cold compression \citep{Albers2022a}. In contrast, along the Hugoniot explored here, only the B1 structure and melting is expected, and our lowest reported pressure is already beyond the stability region of the rB1 phase. Nonetheless, to minimize the possible structural influences, we also evaluated the IAD restricted to the K$\beta$$^{\prime}$ satellite region (IAD K$\beta$$^{\prime}$ satellite in Fig.~\ref{Fig-spin crossover}). To do so, we first aligned the XES spectra to the main K$\beta$$_{1,3}$ peak and integrated the absolute difference over the satellite region to capture changes specific to the satellite feature, which has been demonstrated to directly track the Fe spin moment, as its total loss of intensity indicates full LS Fe \citep{Badro1999}. Thus, the K$\beta$$^{\prime}$ satellite only-IAD provides an additional, more selective indicator of spin crossover, removing potential pressure-induced effects. In Fig.~\ref{Fig-spin crossover}, both IAD evaluations show a consistent, continuous trend with pressure, indicating a progressive increase of the LS fraction with pressure along the Hugoniot. The agreement between both the full IAD and the satellite only-IAD confirms that the observed spectra evolution is dominated by change in the spin state rather than structural effects.

\begin{figure}[b]
\centering
\includegraphics[width=0.6\textwidth]{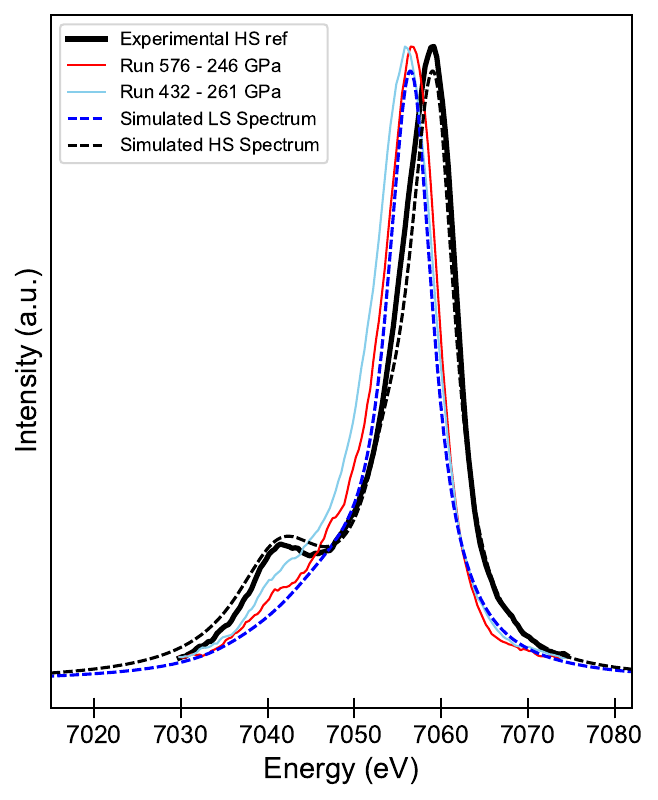}
\caption{Comparison of the experimental spectra with simulated high-spin (HS, black dashed) and low-spin (LS, blue dashed) Fe emission spectra shows. Shocked spectra at 246–261 GPa closely matches the LS simulation, indicating that Fe in FeO approaches a fully low-spin state under these conditions.}\label{Fig-XES simulation}
\end{figure}

At the highest pressures reached in this study, the systematic shift of the K$\beta$$_{1,3}$ main peak together with the near-complete disappearance of the K$\beta$$^{\prime}$ satellite peak (Fig.~\ref{Fig-XES}) indicate that Fe approaches a full LS state \citep{Badro1999}. In the absence of an experimentally measured LS FeO spectrum under comparable conditions, we simulated Fe XES spectra using the code Quanty \citep{Haverkort2016}, allowing a direct visual comparison with our experimental data (Fig.~\ref{Fig-XES simulation}). Details about the simulations can be found in the Supplementary Materials. Two of spectra (e.g., \#576 at 246~GPa and \#288 at 206~GPa) show near-complete loss of the K$\beta$$^{\prime}$ feature, whereas our highest pressure spectrum (432 at 261~GPa) exhibits the furthest energy shift of the K$\beta$$_{1,3}$ main peak but still retains a discernible K$\beta$$^{\prime}$ shoulder (Fig.~\ref{Fig-XES},~\ref{Fig-XES simulation} and~S4). Comparison with the LS simulated spectra clearly shows the residual K$\beta$$^{\prime}$ intensity in our highest pressure spectra (Fig.~\ref{Fig-XES simulation}). This comparison reinforces that our experimental spectra follow a continuous spin crossover, with the fraction of LS Fe increasing progressively with pressure and reaching a full LS state at $>$260~GPa. Calculation of the full IAD using the simulated HS and LS spectra yields a value of 0.43, which can be taken as a quantitative reference for full LS state. This suggests that FeO under CMB conditions contains approximately 50 to 55\% of LS Fe.

Our results significantly extend the pressure-temperature conditions of the measured spin transition in FeO. Previous cold-compression experiments reported the sluggish nature of the spin crossover at ambient temperature which begins at $\sim$70-80~GPa in the rB1 phase and potentially reaching a full LS state between $\sim$120 and 200~GPa in the B8 phase \citep{Greenberg2023, Albers2022a,Hamada2016, Ozawa2011Spin,Mattila2007,Badro1999,Pasternak1997}. 


Recently, \citet{Greenberg2023}, investigated the spin-crossover at high-temperature. Their DFT+DMFT calculations (Fig.~1 in \citet{Greenberg2023}) predict the onset of spin crossover in FeO-B1 at $\sim$70~GPa near 1500~K, shifting to $\sim$125~GPa at around 4000~K, and the full LS state reached at $\sim$100~GPa at 1500~K and $\sim$170~GPa at 4000~K. These results support the persistence of HS FeO-B1 at CMB conditions. Similarly, our XES measurements along the Hugoniot also suggest the persistence of a mixed-spin state in FeO, with HS iron still present at pressures relevant to the CMB ($\sim$135~GPa) at high temperature. In contrast, \citet{Greenberg2023} experimental XES results (Fig.~5 in \citet{Greenberg2023}) suggest a different trend, with a mixed-spin state ($\sim$60\% HS fraction) at 110~GPa and 1650~K but do not extend to higher pressures relevant to the CMB, leaving the calculations untested experimentally. Our XES measurements along the Hugoniot directly addressed this gap, measuring a LS fraction of 50 to 55\% at at CMB conditions ($\sim$135~GPa). A qualitative comparison with the XES spectra of \citet{Greenberg2023} is provided in supplementary material, (Fig. S6), together with a discussion of the limitations.

The divergence between calculated and experimental results in \citet{Greenberg2023} study likely reflects technical challenges inherent to LH-DAC experiments: the small scattering volume requires long acquisition times (minutes to tens of minutes per spectrum), stable laser heating, and precise alignment of the probed and heated areas to minimize the effect of thermal gradients. In addition, the XES signal is collected at 90° through an X-ray-transparent Be gasket, probing the entire sample across pressure —and, when heated, temperature—gradients, potentially obscuring the LS signature. 
Together, these factors complicate high-temperature XES measurements, explaining their limited availability \citep{Greenberg2023, Kaa2022,Albers2023, Albers2022, Albers2022a}.

The role of temperature is central in the spin transition, as elevated temperatures generally shift the spin transition toward higher pressures. Accordingly, in our shock compression experiments, the spin state is likely influenced by the extreme temperatures reached along the Hugoniot. A comparable temperature-dependent trend has been reported in ferropericlase, where the spin crossover occurs over a narrower pressure range at room temperature but broadens and shifts to higher pressures at high temperature (e.g., \citep{Trautner2024,Sun2022,Yang2021,Wu2009,Tsuchiya2006}). Our XES spectra suggest that a full LS state is nearly achieved above 240~GPa along the Hugoniot, suggesting that at lower temperatures, such as those expected at the Earth's outer-core, a full LS state is likely achieved at similar pressure. However, the experimental results of \citet{Greenberg2023} show a non-uniform temperature effect on the spin state: below 80~GPa, higher temperatures increase the LS fraction, whereas at 90~GPa the opposite trend is observed, with $\sim$67\% HS-Fe at $\sim$1800~K and a fully HS state at $\sim$2050~K. This discrepancy may reflect the experimental challenges of probing XES at high temperatures, as discussed earlier, or illustrate a more complex interplay of pressure, temperature and electronic state.

Such electronic temperature-dependent behavior has also been suggested in the recent theoretical work by \citet{Ho2024}. They demonstrated that the IMT in FeO is temperature-dependent. According to their eDMFT calculations, at low temperature (T $\leq$ $\sim$370K) FeO undergoes a sharp transition from Mott insulator to metal. In contrast, at high temperature ($>$2000~K) the IMT is not direct, FeO exists in an intermediate quantum-critical state, where the e$_g$ gap closes only at higher compression, which drives metallization and simultaneously initiates the spin crossover at $\sim$140 to 150~GPa. However, at these pressures, our experimental results already attest to $\sim$50 to 60\% of LS Fe.

Even though the degree of coupling between these transitions remains uncertain, the continuous spin crossover observed here is robustly supported by our data and provides a foundation for considering its broader geophysical consequences. The HS to LS transition, through its 20 to 40\% reduction in iron ionic volume, directly affects FeO's bulk modulus and thus produces abrupt changes in the seismic wave velocity. Similar to ferropericlase, where the spin transition modifies the bulk modulus, the crossover in FeO has the potential to generate depth-dependent seismic anomalies such as those observed at the CMB \citep{Trautner2023, Lin2013}. However, the broad transition interval observed in this study, relevant to much of the Earth’s mantle and core, also suggests that the spin-state-related effects occur progressively with depth. Consequently, the geophysical and geochemical impacts might be distributed over a wide pressure range, rather than localized to a narrow depth range in the Earth's mantle.


Although direct links between the FeO end-member spin state and mantle mineral properties are not straightforward, our results imply that HS iron would persist in FeO-rich melts trapped at the base of the lower mantle at the end of the mantle crystallization \citep{Boukare2025,Boukare2015}. Changes in spin state may also affect iron partitioning between metal and silicate phases and/or between solid and liquid, potentially influencing both mantle and core evolution. Continued efforts to connect electronic transitions to mantle dynamics, phase equilibria, and transport properties are essential to improve geodynamic models. Moreover, 
the spin-state behavior reported here provides an important benchmark for future ab initio calculations. In particular, constraining the role of FeO in Fe–FeO liquid mixing \citep{Komabayashi2014} and its associated electronic properties will be key to fully assess its impact on deep Earth's core processes.


\section*{Methods}\label{Methods}

\subsection*{Sample characterization}\label{Sample}

The \ce{Fe_{1-x}O} sample was purchased from Mateck Material and characterized by single crystal XRD X-ray diffraction at the Institute of Mineralogy, Physics of Materials, and Cosmochemistry (IMPMC) in Paris. The \ce{Fe_{x}O} (x~=~0.9) stoichiometry was determined following the relationship established in \cite{McCammon1984}. Refined lattice parameters are provided in Supplementary Table S1. The FeO targets were between 40 and 70~$\mu$m thick.

\subsection*{Laser-driven shock compression experiments}\label{Experiments methods}

\subsubsection*{LULI2000 experiments} 
\begin{figure}[b]
\includegraphics[width=\textwidth]{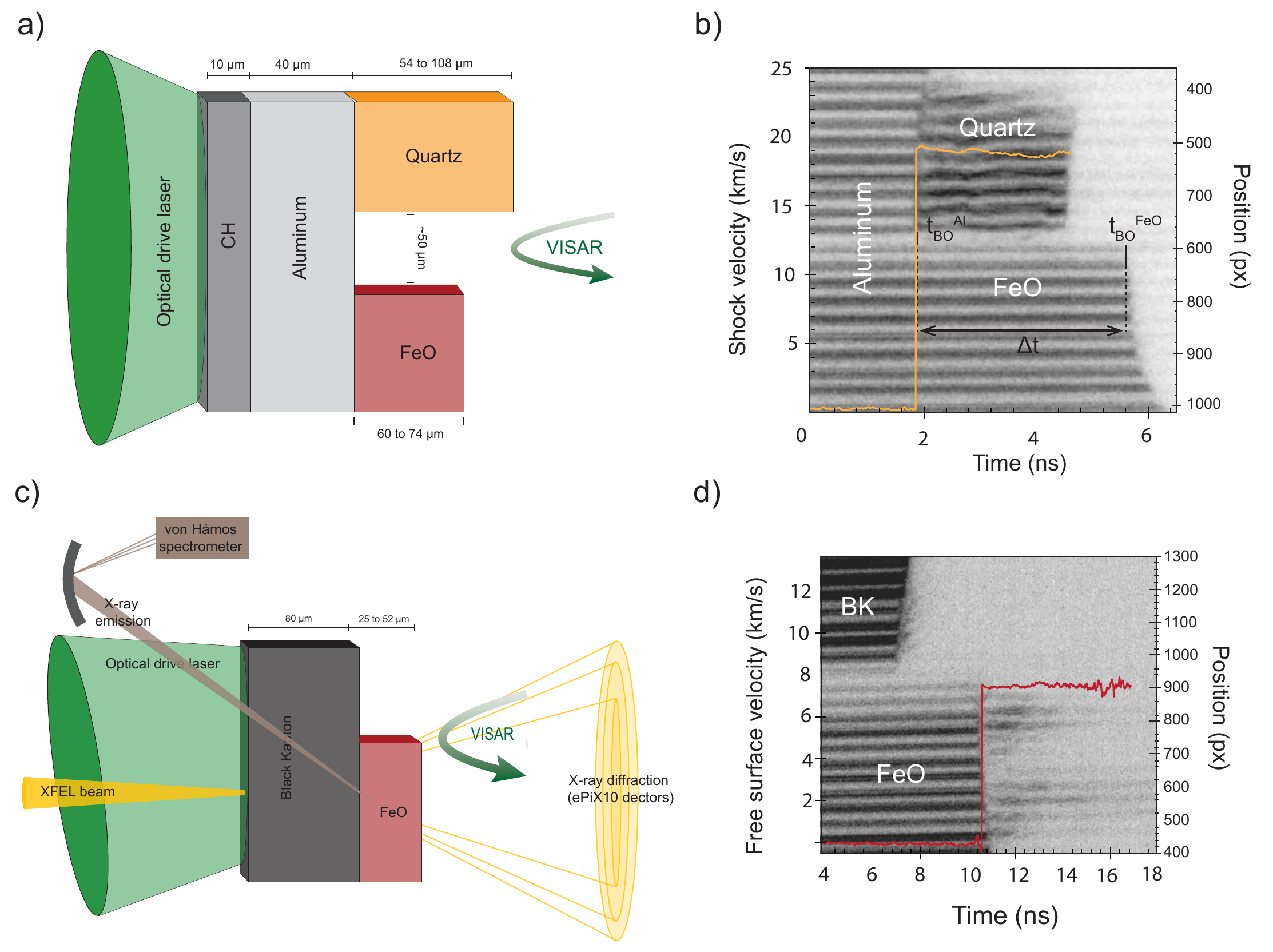}
\caption{Schematic experimental setup: a) Target designs for FeO Hugoniot measurements at LULI2000. b) VISAR record from shot 2 using target design from a) with the extracted shock velocity profiles superposed on the VISAR image. The shock velocity in quartz is determined from the shift in the fringe pattern. FeO is opaque, resulting in no fringe motion being observed in the bottom part of the VISAR image. The mean shock velocity ($\overline{U}$$_{S}$$^{FeO}$) from the shock transit time in FeO ($\Delta$t) calculated by measuring the difference in the breakout times between Al (t$_{BO}^{Al}$) and FeO (t$_{BO}^{FeO}$). c) Schematic representation of the experimental setup geometry at the MEC end station. d)  VISAR record from shot 279 using target design from c) with the extracted free-surface velocity profiles superposed on the VISAR image.}
\label{Experimental Design}
\end{figure}

Flat-top pulses (5 or 10~ns duration) of the high-energy drive laser were focus on the target to a 500~$\mu$m-diameter focal spot using a continuous phase plate. Target consisted of a 10~$\mu$m-thick plastic (CH) ablator and a 40~$\mu$m-thick aluminum layer, which served to reduce pre-heating and act as a pusher to generate a uniform shock wave. Mounted onto the aluminum layer in a side-by-side configuration were an $\alpha$-quartz (54 to 108~$\mu$m thick, $\rho$$_{0}$=2.65~g/cm$^{3}$) and the FeO sample (60 to 74~$\mu$m thick), separated by a $\sim$50 $\mu$m gap (Fig \ref{Experimental Design}a and b). Because FeO is opaque, its thermodynamic conditions during the shock compression were determined by impedance matching technique \citep{Brygoo2015, zeldovich1966}, using the $\alpha$-quartz and aluminum layer used as reference materials \citep{Desjarlais2017,Knudson2013,Hicks2005}. Under shock compression, a reflecting shock front is generated in the $\alpha$-quartz, allowing direct measurement of the shock velocity in quartz (U$_{S}$$^{quartz}$) at the Al/quartz interface using the time-resolved velocimetry diagnostic VISAR (Velocity Interferometer System for Any Reflector)\citep{Barker1965,Barker1970,Barker1972}. The thermodynamic conditions in the aluminum layer are given by the intersection between the quartz Hugoniot at known U$_{P}$$^{quartz}$ and the reverse Hugoniot of Al as an approximation of its release isentrope (Fig. S1). Once the thermodynamic state of aluminum was established, we applied the impedance matching at the aluminum/FeO interface. Since FeO has a higher impedance than aluminum, the shock wave propagating from the aluminum into the FeO sample is partially reflected at the interface and propagated backward, reshocking the aluminum. The conditions in FeO are then given by the intersection between the reversed Hugoniot of Al and the Rayleigh line of FeO. The Rayleigh line is constructed using the initial density and the mean shock velocity in FeO (($\overline{U}$$_{S}$$^{FeO}$)), which is determined from the sample thickness and the shock transit time through the FeO layer. The transit time ($\Delta$t) is obtained by measuring the difference between the shock breakout times in the Al and FeO layers (Fig.~\ref{Experimental Design}b). To ensure reliable measurements of the mean shock velocity, each target was individually characterized and its thickness accurately measured. In addition,  both the target design and laser pulse were optimized to produce a steady shock. Shock stationarity was monitored using a quartz witness positioned next to the sample (Fig.~\ref{Experimental Design}a), in which the instantaneous shock velocity was measured. As shown in (Fig.~\ref{Experimental Design}b), the shock remains highly stationary in the quartz witness. This observation, together with the planarity of the shock front, provides confidence in the reliability of the mean shock velocity measurements. A small gap between the sample and the quartz witness enabled direct and unambiguous identification of the shock breakout time from the Al layer. This geometry also allowed the contribution of the glue layer to be assessed and shown to be negligible due to its minimal thickness. Uncertainties were estimated using a Monte Carlo approach (10000 iterations per data point), accounting for uncertainties in the quartz shock velocity, the mean shock velocity in FeO, and the initial sample density. A schematic impedance match construction is provided in the Fig. S1. Here, we report 5 shots from the LULI2000 facility (Table \ref{TAB-Results Hugoniot} and Supplementary Table S3).

\subsubsection*{MEC experiments} 

At the Matter in Extreme Conditions (MEC) end station of LCLS, in situ X-ray diffraction (XRD) and X-ray emission spectroscopy (XES) measurements were performed during the shock compression. The frequency-doubled Nd:glass drive laser delivered flat-top temporal pulses with a duration of 8~ns and a 300~$\mu$m focal spot on the target, shaped using a continuous phase plate. The targets consisted of a 79~$\mu$m-thick polyimide (\ce{C22H10N2O5}, also known as black Kapton) ablator, coated with a 300~nm layer of aluminum, onto which the FeO sample was glued. Two target configurations were used: (\textit{i}) a “full” design, in which the FeO fully covered the underlying layers, or a (\textit{ii}) “step” configuration, which allowed direct observation of the shock breakout in the Black Kapton on the VISAR image (Fig \ref{Experimental Design}d). A total of 13 shots were carried out over laser energies of 15 to 73~J, generating pressure from 82 and 261~GPa on the FeO Hugoniot (Table S2).

The X-ray probing was timed at various delays, ranging from 8 to 15~ns with respect to the optical laser pulse, to ensure that the measurement occurred while the shock front was propagating approximately in the first 20~nm into the FeO sample. The hydrodynamics were monitored with the VISAR system operating at 532 nm. Several approaches were used to determine the pressure and density of the shocked FeO at MEC:

 \begin{itemize}
     
        \item[\textbf{a.)}] For targets with a step design, the shocked properties of FeO were determined by impedance matching from Black Kapton to FeO, using the mean shock velocity in Black Kapton: $\overline{U}_S^{BK}$ = dBK / $\Delta$t where d is the Black Kapton thickness and $\Delta$t the  transit times, together with the SESAME equation of state for parylene (SESAME 7770) evaluated at the density of Black Kapton. This approach was preferred over the direct determination of the $\overline{U}_S^{FeO}$, to account for the limited accuracy of the FeO thickness determination during the first experimental campaign. In contrast, Black Kapton is reproducible and thickness was well characterized prior the first campaign, providing a more reliable determination of the shock FeO conditions.



        \item[\textbf{b.)}] For the targets with a full design or with a step not visible on the VISAR image, i.e., for which the breakout of the Black Kapton could not be read, the transit time could not be used to determine the shocked conditions of FeO.  Therefore, the determination of the FeO pressure relied on a calibration between the Black Kapton’s ablation pressure and the laser drive energy at MEC. For this calibration, the Black Kapton ablation pressure was obtained by measuring its mean shock velocity ($\overline{U}_S^{BK}$)  from transit times, following the same procedure as in step a.). The dataset comprised all the shots from the same LCLS campaign (L10020) that employed the same phase plate (300) and pulse duration (8ns), where the breakout of Black Kapton was clearly visible in the VISAR images. In total, 81 shots were used for this calibration. Once the $\overline{U}_S^{BK}$ is obtained, the Black Kapton particle velocity and pressure (${U}_P^{BK}$ and P${P}_{ablation}^{BK}$) can be calculated using its equation of state. The impedance matching is then applied for the determination of the FeO shocked state. A schematic representation of the impedance match construction is provided in Fig. S3. The laser drive energy vs Black Kapton ablation pressure calibration is plotted in Fig.~S4.

        \item[\textbf{c.)}] For a couple of targets (runs  441 and 444), a LiF window was glued onto the FeO sample to monitor the FeO/LiF interface velocity. These targets were shot during our first campaign, for which the FeO sample thickness could not be determined with sufficient accuracy. As a result, we chose not to use the LiF measurements to track the particle velocity, as the thickness uncertainty prevented a reliable assessment of shock decay. We emphasize that such release effects were avoided in the X-ray measurements by probing the sample sufficiently early during shock propagation.
        

        \item[\textbf{d.)}] Lastly, for the low-pressure shots below melting—where reflectivity in the sample was still maintained—the particle velocity (U$_{P}$$^{FeO}$) was inferred from the free-surface velocity (U$_{fs}$$^{FeO}$) of the FeO, using the approximation  U$_{fs}$ = 2$_{P}$, valid at moderate pressures \citep{zeldovich1966}. Once the FeO shocked states were determined from U$_{S}$$^{FeO}$, U$_{P}$$^{FeO}$, and $\rho$$_{0}$$^{FeO}$. This procedure was applied to one shot (\#279 - Fig.~\ref{Experimental Design}d) conducted during our second MEC campaign, and for which the sample thickness was better characterized than during the first campaign. Because both U$_{P}$$^{FeO}$ and U$_{S}$$^{FeO}$ were directly measured for this shot, from free-surface velocity and transit time respectively, the pressure was determined using the Rankine-Hugoniot relations without relying on impedance matching, providing an absolute measurement of the FeO shock state. This shot was therefore included in the Hugoniot determination (Fig.~\ref{Fig-Hugoniot}).
    
 \end{itemize}

The uncertainties from the impedance matching were propagated using a Monte Carlo approach (10000 iterations per data point), accounting for uncertainties in the Black Kapton pressure, the mean shock velocity in FeO, and the initial densities of all materials. Pressure was obtain by impedance matching between Black Kapton and FeO, for U$_{P}$ $>$ 3.8 km.s$^{-1}$, the Black Kapton reshock state was calculated using the symmetric Hugoniot derived from the SESAME equation of state for parylene (table \#7770). Detailed extended data can be found in Table S2.

\subsection*{X-ray diagnostics}
During the shock compression, the samples were probed by a quasi-monochromatic incident X-ray beam set at either 8, 9, or 17~keV 
with a temporal pulse length of $\sim$60~fs focused on the target to a spot size of $\sim$30~$\mu$m diameter FWHM. The X-ray spot size was smaller than the compressed layer region, which limited the pressure and temperature gradients in the probed area in the sample. A reference measurement was collected prior to each shot, then the target was moved to a fresh spot for the real shot.  

\textbf{X-ray diffraction (XRD):} The XRD signal was collected in the transmission geometry at 9 and 17~keV using four ePix10k detectors (Fig.~\ref{Experimental Design}c) \citep{Blaj2015,Carini2016}. No self-absorption corrections were applied to the XRD images. At this energy, the XRD setup covered an angular range of 9°~to 75°~2$\theta$, corresponding to a Q-range of 14 to 82 nm$^{-1}$. The Q0 (2$\theta$: 15 to 35, Q-range: 22 to 50~nm$^{-1}$) and Q1 (2$\theta$: 9 to 28, Q-range: 13 to 40~nm$^{-1}$) detectors were used for the analysis. The two-dimensional (2D) diffraction patterns were integrated using the Dioptas software \citep{Prescher2015}, including sample-detector distance, polarization, and tilting, calibrated with \ce{CeO2}.

\textbf{X-ray emission spectroscopy (XES):} XES signals were collected backward, on the drive laser side of the target (Fig.~\ref{Experimental Design}c), at 8 and 9~keV by a multicrystal energy dispersive spectrometer on a von Hamos geometry \citep{AlonsoMori2012}. The details of the setup can be found in \citet{Shim2023}.

\backmatter

\bmhead{Data availability}
The raw XRD, XES, and VISAR images used in this study will be made available on a Figshare repository after acceptance of the paper.
The data analysis code for XES is available at https://github.com/Xuehui-Wei/IXE. 

\bmhead{Supplementary information}

\bmhead{Acknowledgments}
This work was supported by the ANR grant MIN-DIXI (ANR-22-CE49-0006). Dynamic compression experiments were performed at the MEC instrument of LCLS, supported by the US Department of Energy (DOE) Office of Science, Fusion Energy Science under contract SF00515 and FWP 100182 were supported by LCLS, a National User Facility operated by Stanford University on behalf of the US DOE, Office of Basic Energy Sciences. We thank the staff of LULI2000 for their support in conducting the experiments and Frederic Lefevre for target assembly. G.M., A.R., and L.L. would like acknowledge support from the CRNS travel grant GoToXFEL. L.L. and G.M. thank H. Elnaggar for performing the HS–LS spin simulations, as well as S. Merkel and J. Chantel from the UMET laboratory (Lille) their support with target preparation. L.L. also thanks Ian Ocampo for fruitful discussions.

\bigskip



\bibliography{biblio}

\end{document}


 
\section*{Supplementary information for:}
\vspace{7mm}
\centering\textbf{\Large Spin crossover in FeO under shock compression}

\vspace{7mm}

Lélia Libon, Alessandra Ravasio, Silvia Pandolfi, Yanyao Zhang, Xuehui Wei, Jean-Alexis Hernandez, Hong Yang, Amanda J. Chen, Tommaso Vinci, Alessandra Benuzzi-Mounaix, Clemens Prescher, François Soubiran, Hae Ja Lee, Eric Galtier, Nick Czapla, Wendy L. Mao, Arianna E. Gleason, Sang Heon Shim, Roberto Alonso-Mori, and Guillaume Morard.

\vspace{7mm}

\textbf{Corresponding authors:}\\
llibon@carnegiescience.edu\\
alessandra.ravasio@polytechnique.edu\\

\justifying
\clearpage  

\section{Sample characterization}\label{Sample}

The \ce{Fe_{x}O} sample was purchased from Mateck Material and characterized by single crystal XRD X-ray diffraction at the Institute of Mineralogy, Physics of Materials, and Cosmochemistry (IMPMC) in Paris. The \ce{Fe_{x}O} (x~=~0.9) stoichiometry was determined following the relationship established in \citet{McCammon1984}. The results from the X-ray diffraction measurements, including the refined lattice parameters, are provided in supplementary Table S1. The FeO targets were between  40 and 70~$\mu$m thick.

\section{Schematic representation of the impedance match construction of shots conducted at LULI2000} \label{LULI2000}

\begin{figure}[h]
\centering
\includegraphics[width=0.9\textwidth]{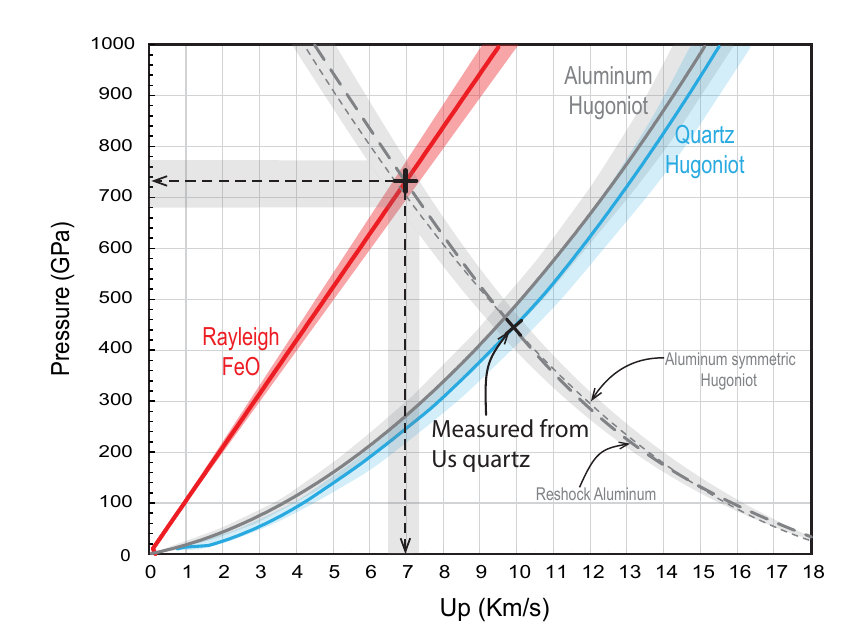}
\caption{Schematic representation of the impedance match construction of shots conducted at LULI2000, using $\alpha$-quartz and aluminum as the reference material for the determination of the Hugoniot state of FeO.}
\end{figure}

\clearpage

\section{Porosity correction of \cite{Jeanloz1980}'s data} \label{PorosityCorrection}

\begin{wrapfigure}{l}{0.5\textwidth}
\includegraphics[width=0.9\linewidth]{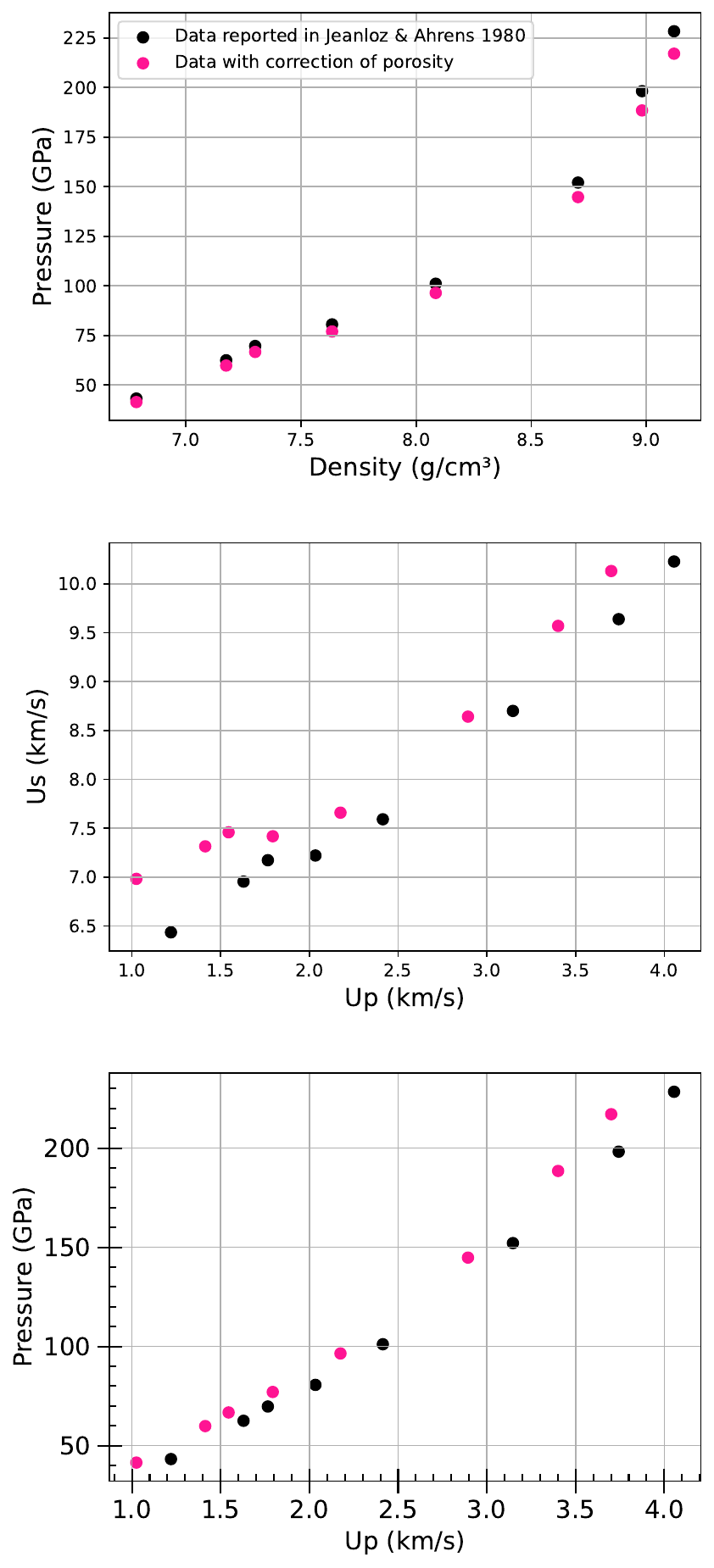} 
\caption{Porosity correction applied to FeO Hugoniot data from \cite{Jeanloz1980}. Black: Original data and red: after correction of porosity.}
\label{Fig-PorosityCorrection}
\end{wrapfigure}

The FeO targets from \cite{Jeanloz1980}'s earlier Hugoniot study were hot-pressed polycrystals containing $\sim$3\% porosity. To account for this porosity and to provide Hugoniot relationships that include both previous and new data, we applied a porosity correction to their dataset.

To do so, we also re-evaluate the stoichiometry coefficient and crystal density of their sample. \cite{Jeanloz1980} reported a lattice constant of 4.308 $\pm$ 0.4~$\mathring{A}$ and used \cite{hentschel1970} to calculate the stoichiometry coefficient and crystal density (x~=~0.94 and $\rho$$_0$$^{crystal}$ = 5.731 $\pm$ 0.4~g/cm$^3$). We recalculated their stoichiometry coefficient and crystal density following \cite{McCammon1984}, obtaining x~=~0.946 and $\rho$$_0$$^{crystal}$ = 5.79 g/cm$^3$. To correct for porosity, we used the equations from \cite{vassiliou1981}, along with $\rho$$_0$$^{porous}$ = 5.502~$\pm$~0.006~g/cm$^3$, the recalculated $\rho$$_0$$^{crystal}$, and the ambient Grüneisen parameter ($\gamma$$_0$ = 1.41~$\pm$~0.05) from \cite{Fischer2011EOS}. The corrected data are plotted alongside those reported in \cite{Jeanloz1980} in Fig.~S\ref{Fig-PorosityCorrection}.

Later, \cite{Yagi1988} re-investigated the FeO hugoniot from 63 to 121~GPa, using a specifically pore-free sample. Despite the differences in porosity and stoichiometry between the two studies, they reported results very consistent with those of \cite{Jeanloz1980}, observing the same discontinuous increase in density at around 70 GPa. 
 

\clearpage

\section{Schematic representation of the impedance match construction of shots conducted at MEC}

\begin{figure}[h]
\centering
\includegraphics[width=0.9\textwidth]{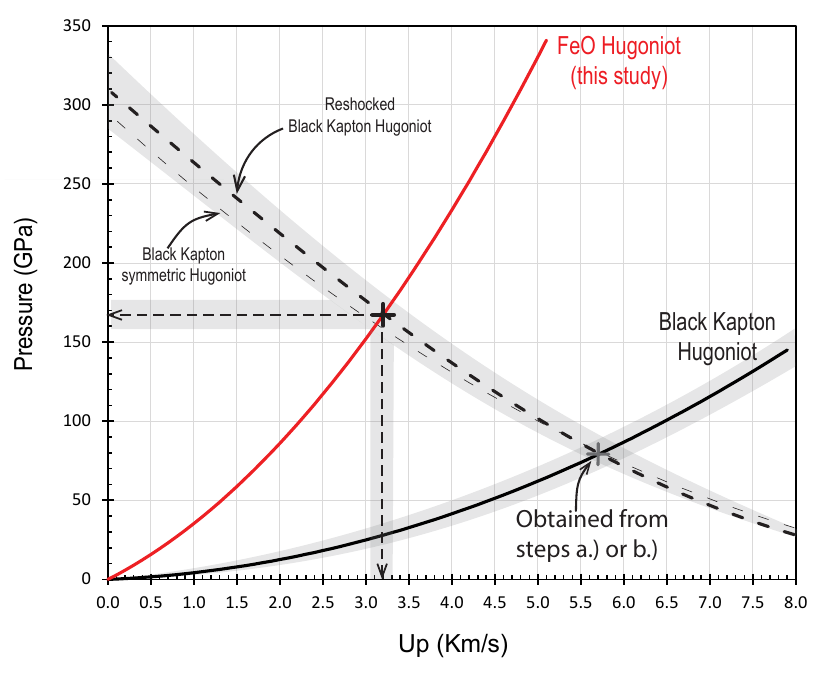}
\caption{Schematic representation of the impedance match construction of shots conducted at MEC, using the the Black Kapton.}
\label{Fig-impedanceMEC}
\end{figure}

\clearpage
\section{Laser drive energy vs Black Kapton ablation pressure}

At MEC, the targets in the “full” design, prevented observation of the Black Kapton breakout on the VISAR, making it impossible to determine the mean shock velocity of the Black Kapton ($\overline{U}_S^{BK}$).
 
\begin{figure}[h]
\centering
\includegraphics[width=0.8\textwidth]{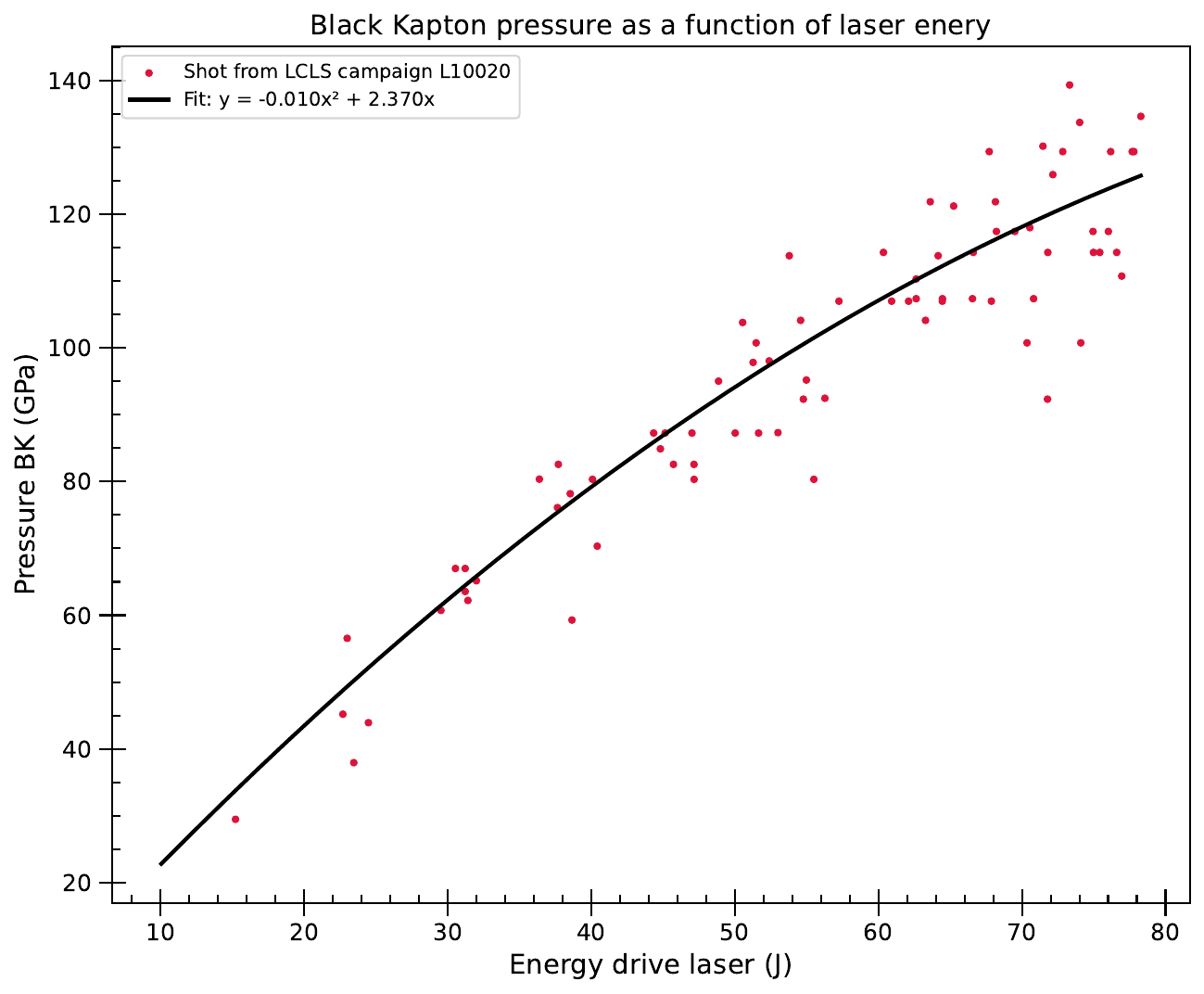}
\caption{Black Kapton pressure as a function of laser enery for the 300~$\mu$m phase plate at the MEC end-station.}
\label{Black Kapton pressure}
\end{figure}

To determine the pressure conditions in the Black Kapton and then in the FeO sample via the impedance matching technique, we relied on a empirical calibration between the laser drive energy and ablation pressure of Black Kapton (Fig.~\ref{Black Kapton pressure}). For this calibration, we used all shots from the same LCLS campaign (L10020) that employed the same phase plate (300) and pulse duration (8~ns), in which the breakout of Black Kapton was clearly visible on the VISAR image.

 The ablation pressure of Black Kapton was calculated from its mean shock velocity ($\overline{U}_S^{BK}$), determined from the shock transit time through the (80~$\mu$m) Black Kapton layer. The transit time was obtained using the breakout timing (t$^{BK}$) assuming that the shock break-in was synchronized with the laser drive $T_0$. 

 Once the ablation pressure was obtained, we applied the impedance matching technique taking, accounting the reshock of the Black Kapton from the FeO sample and combined with Monte Carlos simulations to evaluate the shock conditions in FeO (Fig.~S\ref{Fig-impedanceMEC}). A total of 81 shots were used for this calibration.

\clearpage 

\section{XES spectra}

\begin{figure}[h]
\centering
\caption{Fe K$\beta$ and K$\beta$'emission spectra for all shots. The black curve is the high-spin reference spectra at ambient pressure and the red curve is during the shock. Spectra on the right focus on the K$\beta$' region, after the main K$\beta$ peaks (reference and shocked) has been centered. The difference between the two spectra is graphically represented by the solid grey area.}
\includegraphics[width=0.7\textwidth]{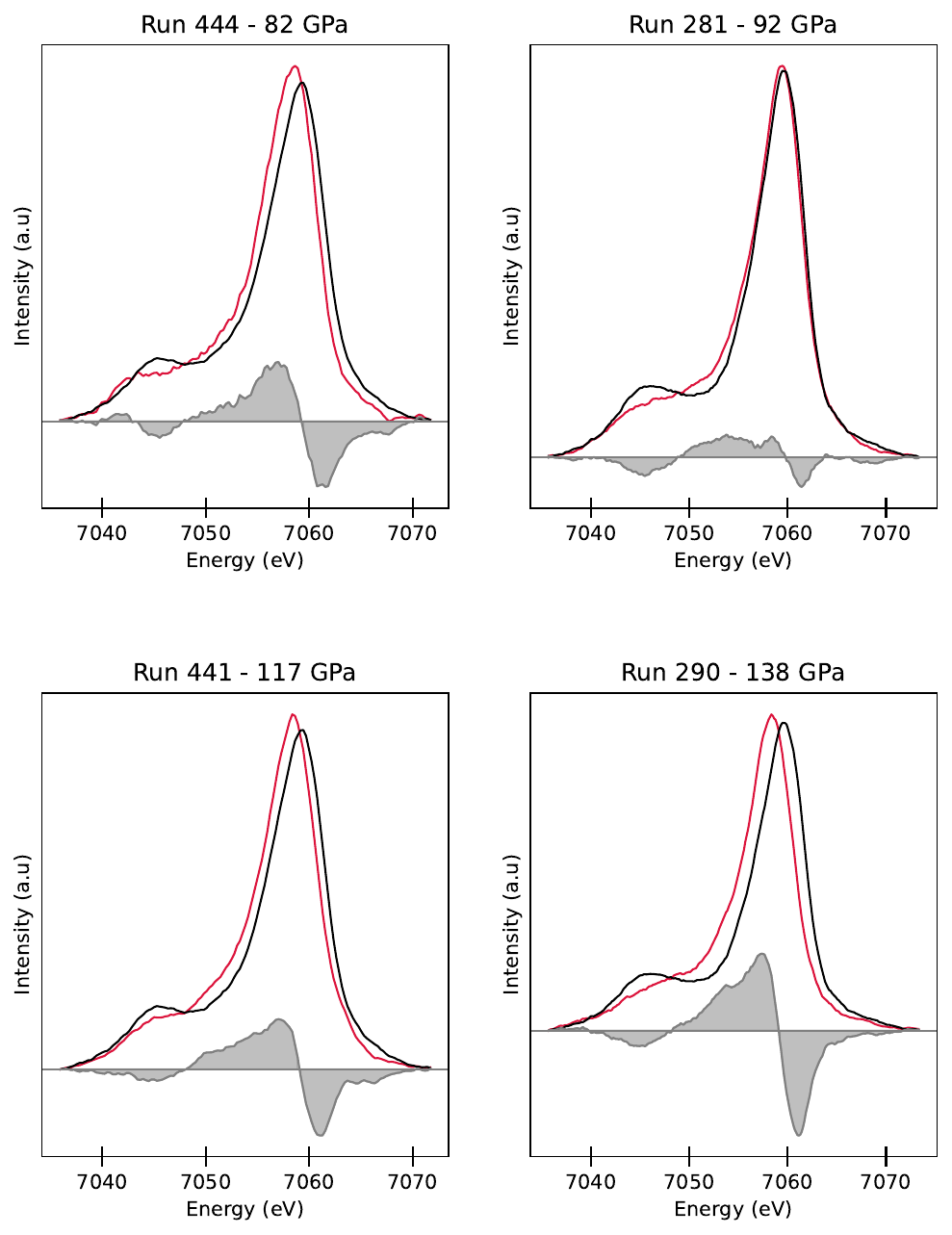}
\end{figure}

\begin{figure}[h]
\centering
\includegraphics[width=0.7\textwidth]{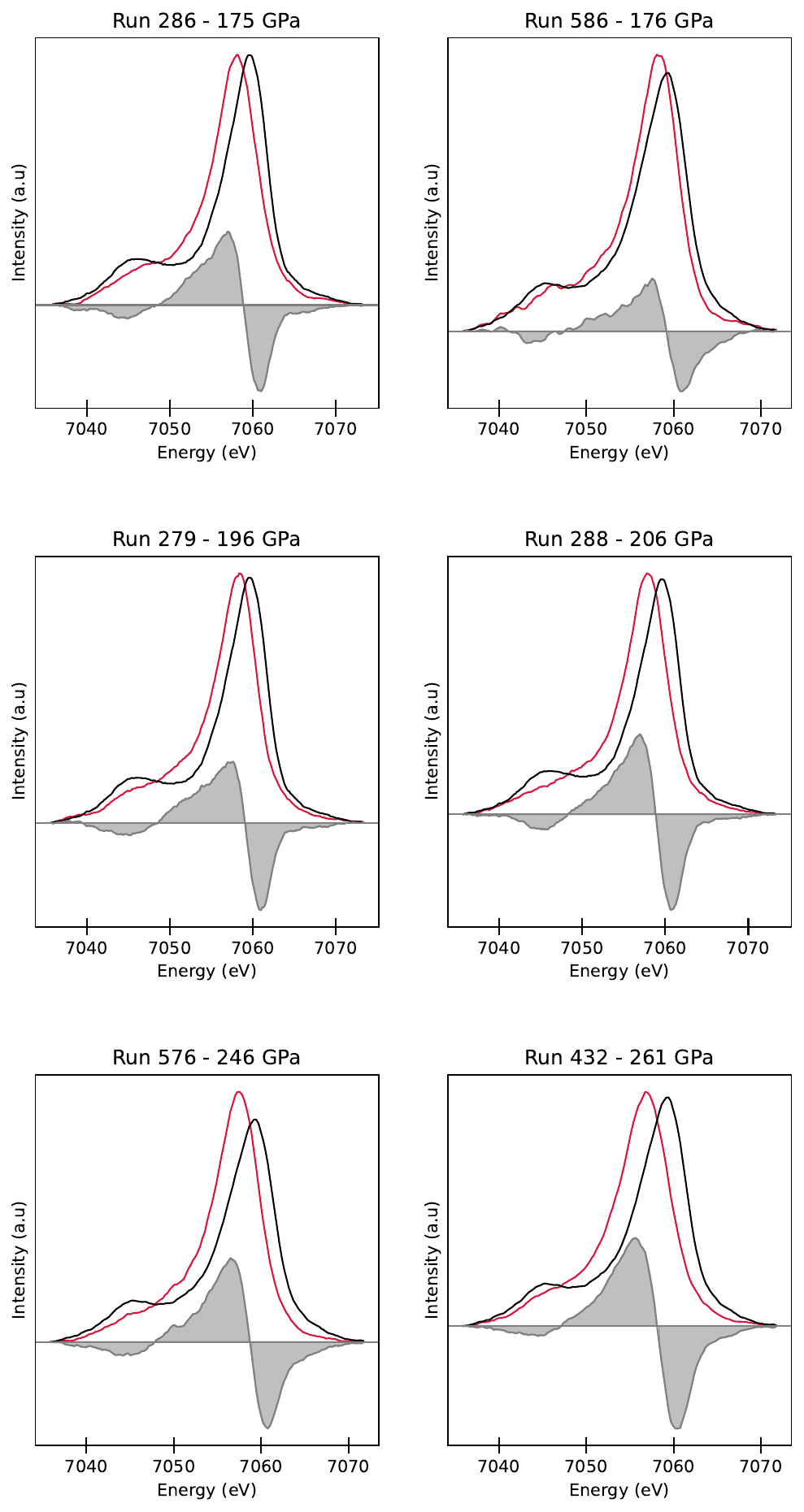}

\end{figure}

\clearpage 

\section{Computational details}

We used a two-step approach to calculate the XES spectra. We started with the thermally populated states corresponding to the 3d$^6$ electronic configuration. A e 1s electron is then excited into the continuum. The intermediate states formed after photo-ionization belonging to the 1s$^1$3d$^5$ electronic configuration are then allowed to decay through a 3p to 1s dipolar electronic transition producing the XES spectra. The Hamiltonian used to describe the system includes contributions among the 1s, 3p, and 3d electrons: i.e., electron-electron, spin-orbit coupling interactions and crystal field splitting. The parameters for an isolated Fe$^{2+}$ ion are given below. \\
\\

Electronic configuration: 3d$^6$ \\
F$^2$$_{dd}$ = 7.675 eV \\
F$^4$$_{dd}$ = 4.770 eV \\
$\zeta$$_d$ = 0.05 eV \\

Electronic configuration: 1s$^1$3d$^6$ \\
F$^2$$_{dd}$ = 7.675 eV \\
F$^4$$_{dd}$ = 4.770 eV \\
$\zeta$$_d$ = 0.05 eV \\
G$^2$$_{sd}$ = 0.07 eV \\

Electronic configuration: 3p$^5$3d$^6$ \\
F$^2$$_{dd}$ = 7.675 eV \\
F$^4$$_{dd}$ = 4.770 eV \\
$\zeta$$_d$ = 0.05 eV \\
F$^4$$_{pd}$ = 9.135 eV \\
G$^1$$_{pd}$ = 11.8041 eV \\
G$^3$$_{pd}$ = 7.1905 eV \\
Z$_p$ = 0.97 eV \\

The high spin state is reproduced using a crystal field splitting (10Dq) of 1 eV while the low spin state is reproduced with a splitting of 3 eV. This is consistent with previous reported literature \citep{Lafuerza2020}.

\clearpage 

\section{Comparison with previous XES measurements}

\begin{wrapfigure}{l}{0.5\textwidth}
\includegraphics[width=0.9\linewidth]{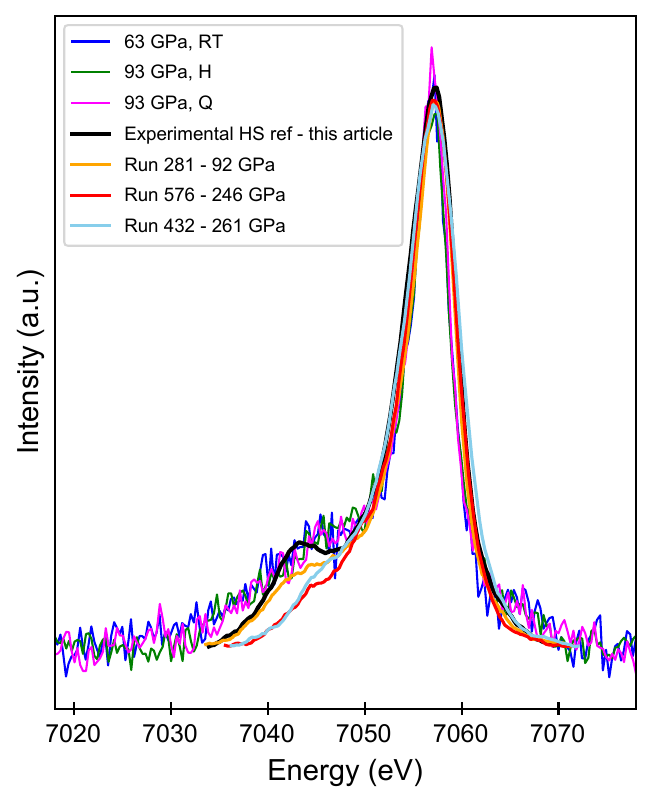} 
\caption{Fe-K$\beta$ X-ray emission spectra collected from FeO from this study and \cite{Greenberg2023}'s study from LH-DAC experiments. The spectra have been centered at the K$\beta$$_{1,3}$ peak around 7058 eV. RT: room temperature, H: hot, and Q: quenched.}
\label{Fig-Greenberg}
\end{wrapfigure}


Several XES spectra were obtain from Fig. 4 of \cite{Greenberg2023}, enabling a qualitative comparison between the two studies. For consistency, all spectra were centered at the K$\beta$$_{1,3}$ peak around 7058 eV. Fig. S6 presents spectra from \cite{Greenberg2023} at 63 GPa and room temperature (RT) and 93 GPa during heating (H) at 2050 K and after quenching (Q), alongside the highest pressure spectra of this study at 246 GPa and 261 GPa at high temperature. Our highest pressure spectra exhibit a greater reducing of K$\beta$$^{\prime}$ satellite intensity, consistent with a predominantly LS state. In contrast, the \cite{Greenberg2023} spectra at 93 GPa the K$\beta$$^{\prime}$ still remain prominent. However, without their highest pressure spectrum at 135 GPa (RT), a complete qualitative comparison is not possible. 

In addition, quantitative, side-by-side comparison of XES spectra from different studies is inherently challenging. In XES spectral features are highly sensitive to instrumental parameters, such as spectrometer geometry, analyzer radius, energy calibration procedure, and X-ray source characteristics which vary between facilities and even between beamtimes at the same beamline. Here in particular, comparison of synchrotron-based measurements with our FEL-based measurements requires caution, as these differences influence both the width of the K$\beta$$_{1,3}$ peak and the relative intensity of the spectrum, strongly limiting comparison between studies. 

\clearpage

\bibliographystyle{cas-model2-names}
\bibliography{biblio} 